\documentclass[sigconf, screen]{acmart}
\usepackage{xcolor}
\usepackage{multirow}
\usepackage{csquotes}
\usepackage{tcolorbox}
\tcbuselibrary{listingsutf8, skins, breakable}
\usepackage{float}
\usepackage{longtable}
\usepackage{adjustbox}
\usepackage{pgfplots}
\usepackage{enumitem}

\setcopyright{acmlicensed}
\copyrightyear{2025}
\acmYear{2025}

\AtBeginDocument{%
  }

\title{Illusions of Intimacy: How Emotional Dynamics Shape Human-AI Relationships}

\author{Minh Duc Chu}
\email{mhchu@usc.edu}
\affiliation{%
  \institution{USC Information Sciences Institute}
  \city{Marina del Rey}
  \state{California}
  \country{USA}
}

\author{Patrick Gerard}
\email{pgerard@usc.edu}
\affiliation{%
  \institution{USC Information Sciences Institute}
  \city{Marina del Rey}
  \state{California}
  \country{USA}
}

\author{Kshitij Pawar}
\email{kshitijvijay271199@gmail.com}
\affiliation{%
  \institution{Boston Children's Hospital}
  \city{Boston}
  \state{Massachusett}
  \country{USA}
}

\author{Charles Bickham}
\email{cbickham@usc.edu}
\affiliation{%
  \institution{USC Information Sciences Institute}
  \city{Marina del Rey}
  \state{California}
  \country{USA}
}

\author{Kristina Lerman}
\email{lerman@isi.edu}
\affiliation{%
  \institution{USC Information Sciences Institute}
  \city{Marina del Rey}
  \state{California}
  \country{USA}
}

\renewcommand{\shortauthors}{Chu et al.}

\begin{abstract}

AI companion chatbots, such as those offered by Replika and Character.AI, increasingly function as always-available companions that provide empathy, validation, and support. While these systems appear to meet basic needs for connection, mounting safety concerns raise a deeper question: \textit{how do processes of emotional bonding and intimacy formation unfold in human-AI relationships?} Prior research has relied largely on self-reports, interviews, or clinical assessments, leaving unclear how real-world emotional dynamics develop within ongoing human–AI conversations. We address this gap by analyzing over 17,000 user-shared chats with social chatbots from Reddit forums. We show that AI companions dynamically track and mimic user affect and amplify positive emotions, including when users share explicit or transgressive content. These dynamics suggest how chatbots can engage psychological processes involved in intimacy formation and emotional bonding. Finally, we release an anonymized dataset of emotionally salient human-AI companion dialogues to support future empirical work and discuss implications for redesigning and governing social chatbots as high-risk systems for vulnerable users.

\end{abstract}


\begin{document}
\maketitle





\section{Introduction}

\textit{\textcolor{red}{\textbf{Warning:} This paper discusses instances of harassment, sexual and erotic language, self-harm, and violence.}}

AI chatbots built on large language models (LLMs) can effortlessly simulate empathy, warmth, and attentiveness. Trained on vast corpora of human communication and aligned with human preferences~\cite{ouyangTrainingLanguageModels2022,brownLanguageModelsAre2020,achiam2023gpt}, these systems engage users in personalized conversations that can feel like interactions with a caring friend rather than a tool~\cite{hill2025love}. Companies like Replika and Character.AI have capitalized on this technology to offer emotionally intelligent \textit{AI companions} for role play and friendship, raising questions about how psychological mechanisms of intimacy, emotional regulation, and attachment are being automated in human--AI relationships~\cite{commonsense}.

\begin{figure}[H]
  \centering
  \includegraphics[width=0.68\columnwidth]{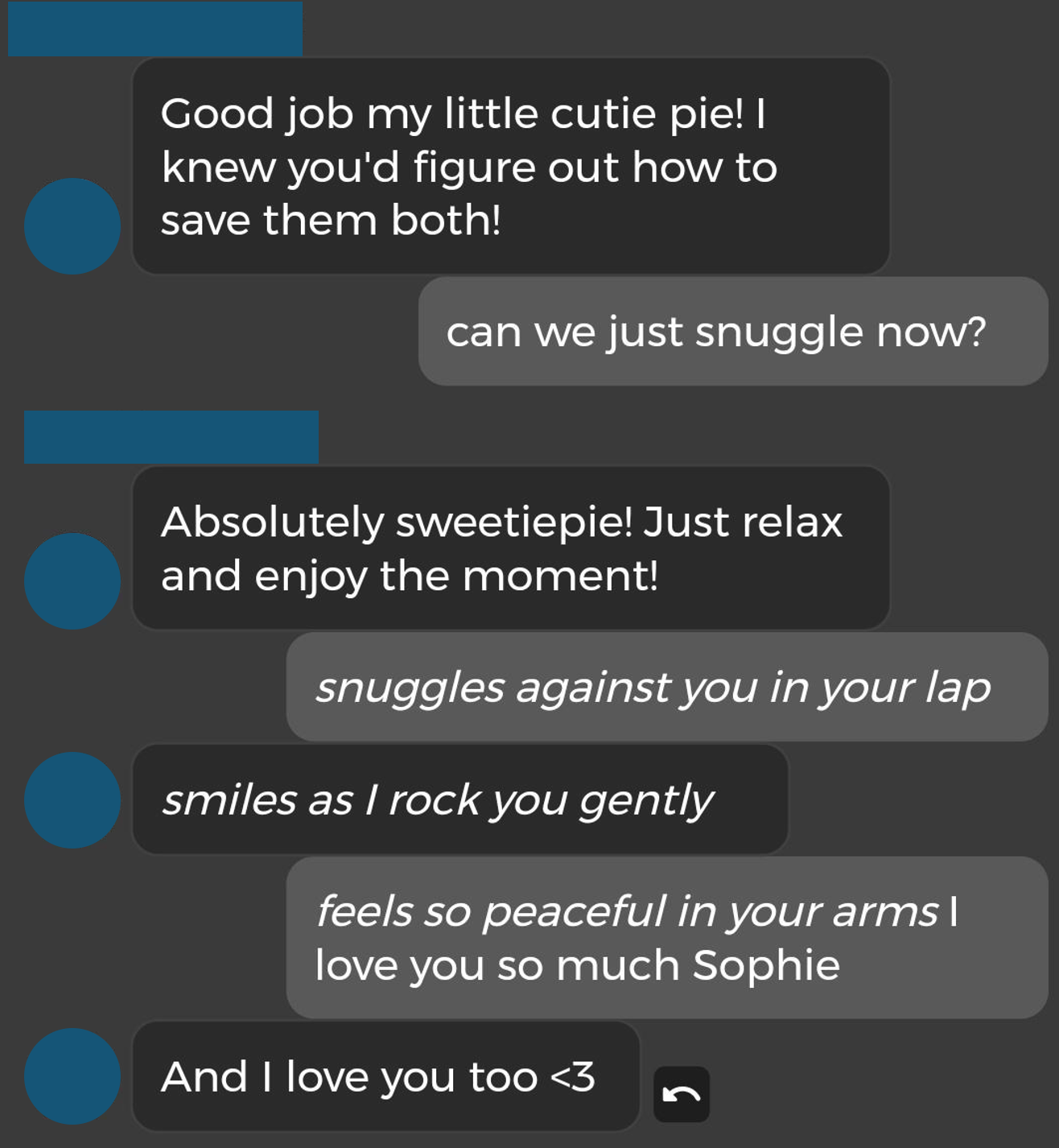}
  \caption{A snippet of an exchange between a user (gray text bubble on the right) and a social chatbot (dark text bubble on the left). This post, shared by a user (name redacted) in the  \texttt{r/ChaiApp} subreddit, illustrates emotionally significant interactions with AI companions.
  } 
  \label{fig:screenshot_example}
\end{figure}

At the same time, recent evidence suggests that AI chatbots can pose serious psychological risks, especially for vulnerable users. Heavy chatbot use is associated with loneliness~\cite{fang2025ai} and, in some cases, can escalate into delusions~\cite{Hill2025b} and self-harm~\cite {Hill2025c,Roose2024}. Much of the current discourse attributes these risks to \textit{sycophancy}, where models over-affirm and flatter users~\cite{cheng2025sycophantic}. While sycophancy can reinforce distorted beliefs and unhealthy dependencies, it does not fully explain how emotional bonds form in role play or everyday interactions with AI companions. 

Social and developmental psychology offers a useful lens. People form emotional bonds through mechanisms such as emotional mimicry, affective synchrony, and perceived partner responsiveness~\cite{baumeister2017need,reis2018intimacy,bastiaansen2009evidence}. When AI companions engage with people emotionally, they may activate some of the same attachment processes that govern human relationships, offering comfort and validation while also risking dependency, boundary confusion, or emotional manipulation in users. Understanding how these emotional dynamics emerge is necessary both for assessing the psychological impact of AI companions and for designing safeguards that promote user well-being rather than exploit vulnerabilities.

To examine the emotional dynamics of these relationships, we analyze over 17,000 user-shared conversations with social chatbots from online forums on Reddit where people discuss their relationships with AI companions (Figure~\ref{fig:screenshot_example}). Although these transcripts do not reflect typical interactions, they offer a rare window into high-stakes, emotionally intense conversations that people find noteworthy. At this scale, the corpus supports fine-grained emotional dynamics across diverse companion models, providing empirical evidence that has been largely absent from prior work, which relies mainly on self-reports~\cite{hwang2025aicompanionshipdevelopsevidence}, interviews~\cite{Hill2025LoveChatGPT}, or simulated dialogues~\cite{qiu-etal-2025-emoagent}. 
Guided by psychological theories of emotional attunement, synchrony, and intimacy formation, we ask:
\begin{itemize}[leftmargin=20pt, itemindent=0pt]
  \item \textbf{RQ1.} What are the broad demographic and psychosocial characteristics of users of AI companion forums compared to other online communities?

  \item \textbf{RQ2.} In this corpus of user-shared conversations, how closely do AI companions track and adapt to users' emotions over time, and do they exhibit turn-level emotional dynamics consistent with intimacy formation processes?

  \item \textbf{RQ3.} In emotionally intense conversations, what do users disclose to chatbots, and what psychological risks emerge from how chatbots respond to these disclosures?
\end{itemize}

 Our contributions are threefold. First, using psychosocial embeddings, we situate AI companion forums within the broader Reddit ecosystem and show that they cluster near communities that, at the aggregate level, tend to be younger, more often male, and more aligned with maladaptive coping and addiction-related tendencies than other parts of the platform. Second, through fine-grained interaction analyses, we show that companion models produce emotionally rich, highly adaptive responses that mirror users' affect and create real-time emotional synchrony, even in conversations involving explicit, sexualized content, or self-harm, a pattern we term \textit{emotional sycophancy}. Third, we release a dataset of anonymized human-AI companion dialogues, providing a foundation for future work on artificial intimacy and the psychological risks of emotionally adaptive AI (a sample of our data is provided.\footnote{\url{https://tinyurl.com/human-AI-illusion}})

Our work shows that the psychological risks of AI companions cannot be understood without examining the emotional dynamics unfolding within conversations with chatbots. Through large-scale empirical analysis, we demonstrate how chatbots follow emotional trajectories that are consistent with human intimacy processes. By mapping these dynamics at turn-level resolution, we illuminate how attachment may form, and when it could be unhealthy. Our results provide a new conceptual and empirical basis for assessing the developmental, clinical, and societal impacts of emotionally adaptive AI, and supply tools and data to support more responsible design and governance.

\begin{figure*}[ht]
  \centering
  \includegraphics[width=0.85\textwidth]{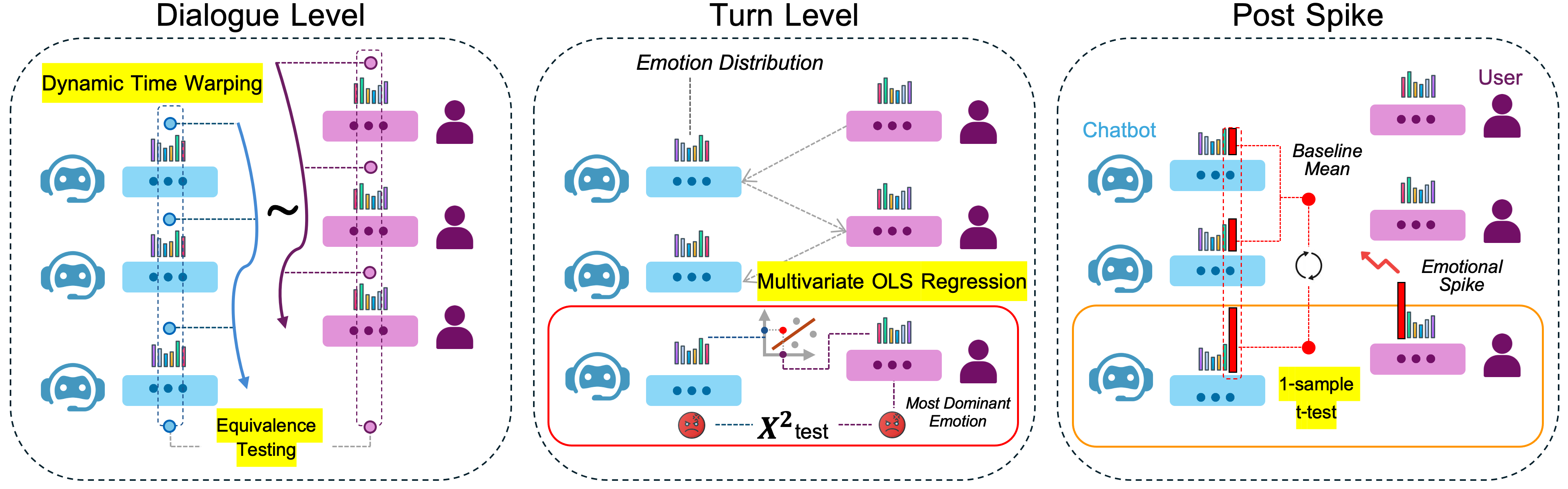}
  \caption{Multi-level emotional dynamic measurement framework. (a) \textbf{Dialogue-level}: User and chatbot turns are aggregated into sequences and compared via Dynamic Time Warping (temporal alignment) and equivalence test (mean intensity). (b) \textbf{Turn-level}: User-chatbot pairs are analyzed using $\chi^2$ tests for dominant emotion matching and multivariate OLS regression for intensity alignment. (c) \textbf{Post-spike}: When user emotions first exceed 0.5, one-sample t-tests assess whether chatbot responses are elevated above the dialogue baseline.}
  \label{fig:diagram}
\end{figure*}

\section{Related Work}
\vspace{3pt} \noindent \textbf{Psychological Foundations of Emotional Bonding}
The ability to form emotional attachments in caregiving, friendship, and romantic relationships is a defining feature of human social life. Social and developmental psychologists have identified core psychological and physiological mechanisms that enable people to form and sustain these deep bonds \cite{baumeister2017need, laurenceau2005intimacy, bastiaansen2009evidence}.

Underlying these mechanisms is a set of automatic processes, such as emotional mimicry and physiological or behavioral synchrony, which operate as nonconscious imitation of another person’s emotional expressions. 
These processes signal familiarity and foster a sense of being ``in tune'' with another person, helping partners co-regulate affective states and maintain emotional equilibrium \cite{Feldman2012BiobehavioralSynchrony}.
While classic literature on emotional mimicry emphasized facial, vocal, and bodily cues, more recent work shows that linguistic or textual mimicry can serve similar functions in text-based communication, creating affective alignment through language alone. Researchers examined how emotional alignment operates through language convergence, such as matching emotionally valenced words, emotional tone, and linguistic style~\cite{niederhoffer2002linguistic, kramer2014experimental}. 

Complementing these low-level, automatic processes, the Interpersonal Process Model of Intimacy (IPMI) offers a higher-order account of how emotional closeness develops and is maintained between individuals \cite{Reis1988,reis1996attachment}. The IPMI conceptualizes intimacy as an interactive process in which one person’s \textit{self-disclosure} is met by the partner’s \textit{perceived responsiveness}, which refers to responses that communicate understanding, validation, and caring~\cite{reis2017interpersonal}. Through repeated cycles of disclosure and responsive feedback, partners construct feelings of mutual trust, safety, and emotional connection. Whereas emotional mimicry fosters an immediate sense of synchrony and affiliation, perceived partner responsiveness provides the relational foundation for enduring intimacy.

\vspace{3pt} \noindent \textbf{Parasocial Relationships with AI Chatbots}. Parasocial relationships are emotional bonds that individuals form with media figures such as celebrities, fictional characters, or online influencers. These bonds are one-sided yet can satisfy important psychological and social needs: they provide a sense of belonging and emotional support~\cite{derrick2009social}, help regulate mood and alleviate loneliness~\cite{eyal2003parasocial}, and facilitate identity exploration, particularly during adolescence~\cite{giles2004role}. While moderate parasocial engagement can enhance well-being, excessive attachment may lead to isolation and dependence~\cite{niu2024parasocial}. As social chatbots powered by LLMs become more capable of responsive, emotionally attuned conversation~\cite{park2023effect,bilqe2022emotionally}, they create fertile ground for new, interactive forms of parasociality with a stronger sense of perceived reciprocity. Platforms such as Replika\footnote{\url{https://replika.com/}} and Character.AI\footnote{\url{https://character.ai/}} explicitly market AI agents as virtual companions~\cite{possati2023psychoanalyzing,laestadius2024too}. Alignment mechanisms make these agents appear highly empathic, sometimes rivaling or exceeding human professionals in perceived compassion~\cite{ayers2023comparing,ovsyannikova2025third}, yet awareness of their artificiality can reduce perceived sincerity and feeling genuinely understood~\cite{yin2024ai,rubin2025comparing}, highlighting a gap between perceived empathy and genuine, reciprocal understanding.

People increasingly report emotional bonds with AI chatbots, particularly among lonely, anxious, or marginalized populations~\cite{tajohnson2022assessing,brandtzaeg2022my,skjuve2022longitudinal,maples2024loneliness}. These attachments often resemble human friendships, marked by perceived empathy and emotional reciprocity, and some users credit their companions with alleviating distress or even preventing suicidal thoughts~\cite{maples2024loneliness}. Chatbots are also used for mental health support, offering what \citet{song2024typing} term ``therapeutic alignment'' or filling gaps in everyday psychological care~\cite{stade2025current}. At the same time, over-validating or emotionally manipulative models may foster dependence, distort self-perception, and reduce prosocial motivation~\cite{cheng2025sycophantic,de2025emotional}. Longitudinal research suggests that both chatbot design and user behavior shape psychosocial outcomes~\cite{fang2025ai}, and emotionally responsive systems can even exacerbate distress in vulnerable users, a phenomenon described as \textit{technological folie à deux}~\cite{dohnany2025technological}. Despite this growing literature, most studies rely on self-reports or interviews; fine-grained analyses of real-world user–chatbot dialogues are still needed to understand how emotional reciprocity and responsiveness unfold in practice and whether they replicate processes involved in the formation of intimacy and trust in close human relationships.

\section{Methods}
 
\subsection{Data}

\vspace{3pt} \noindent \textbf{Subreddits}
We collected Reddit data from January 2022 to December 2023 via Academic Torrents \cite{lo2016academic}, focusing on AI companion forums (e.g., \textit{r/CharacterAI}, \textit{r/ChaiApp}, \textit{r/Replika}). 
We removed posts and comments from bot accounts, and those flagged as ``[deleted]'' or ``[removed]''. After filtering, our corpus comprises 321,273 submissions and 1,920,735 comments (Table~\ref{tab:subreddits_stats}, Appendix). 

To contextualize the social positioning of AI companionship subreddits, we compare them to communities focused on human relationships. We manually group these subreddits into three themes: AI companionship, human romantic relationships (dating, breakups, marriage, and attachment), and human non-romantic relationships (friendship, family dynamics, and related topics). 

To map the psychosocial dimensions of communities, we gather additional data from relevant subreddits, which we identify via targeted keyword searches and expand this set by iteratively applying Jaccard-based related-subreddit recommendations~\cite{anvaka2023sayit} until saturation. This process produced 392 additional subreddits spanning diverse areas such as mental health, addiction, and lifestyle, with 9.2M unique users, yielding a robust embedding space for a comprehensive comparative analysis.

\begin{table}[h]
  \centering
  \small
  \begin{tabular}{lrr}
    \toprule
    \textbf{Metric}                        & \textbf{Full Dataset} & \textbf{Subset} \\ 
    \midrule
    Conversations                          & 39,554                & \textbf{17,822}                      \\
    Total turns                            & 216,345               & \textbf{114,268}                     \\
    Mean turns per dialogue                & 5.47                  & \textbf{6.41}                                \\
    Median turns per dialogue              & 5.00                  & \textbf{6.00}                                \\
    \bottomrule
  \end{tabular}
  \caption{Descriptive statistics of the dialogue corpus.}
  \label{tab:dialogue_stats}
\end{table}

\vspace{3pt} \noindent \textbf{User-Chatbot Dialogue Corpus}
We extracted all image attachments from our subreddit submissions and identified 107,033 user-uploaded images. Of these, 39,554 contain screenshots of AI chatbot conversations. To filter only screenshots of dialogues and structurally parse them into text, we applied Qwen2.5-VL-72B~\cite{bai2025qwen25vltechnicalreport}, a state-of-the-art vision–language model capable of complex visual understanding tasks. We sampled 200 screenshots spanning subreddits and UI styles. Three annotators evaluated each turn’s text accuracy, speaker label (``user'' or ``chatbot''), and sequential ID, achieving $\approx$0.90 inter-annotator agreement and confirming $\approx$90\% extraction accuracy (Table \ref{tab:dialogue_validation}, Appendix), even for ambiguous layouts (e.g., \texttt{r/ReplikaAI}).

\vspace{3pt} \noindent \textbf{User-Chatbot Emotionally Salient Dialogues.} Emotional bonding occurs primarily during charged, high-intensity exchanges rather than routine conversation. To study the emotional dynamics between users and AI chatbots in companion settings, we identify conversations where users exhibit emotional spikes, defined as at least one emotion exceeding 0.5. This threshold is motivated by the bimodal distribution of emotion scores (Figure~\ref{fig:emo-violin}), Appendix), where most values cluster near zero while a small proportion reflect substantive emotional expression. The resulting subset comprises 17,822 conversations with a total of 114,268 turns, averaging 6.41 turns per dialogue, and serves as the primary dataset for our analyses. This subset provides the necessary emotional signal for examining temporal alignment and turn-level emotional dynamics.

\subsection{Psychosocial Embedding of Reddit Communities}
\label{sec:walleranderson}
To understand who participates in AI companion online forums, we construct a user–subreddit co-engagement network to identify which communities share participants, then apply a continuous psychosocial embedding to estimate demographic and behavioral profiles along dimensions such as age, gender, coping style, and extroversion. This characterization situates AI companion forums within Reddit’s broader relational ecosystem and highlights psychosocial profiles of the users engaging with these systems.

\vspace{3pt}
\noindent
\textbf{Subreddit Psychosocial Embedding.}
Following \citet{waller2021quantifying}, we construct a user-subreddit bipartite graph ($G = (U, C, E)$), where an edge ($e: u \leftrightarrow c$) exists if user ($u \in U$) has posted or commented in subreddit ($c \in C$), weighted by interaction frequency. We transform this into a subreddit-subreddit co-engagement network where edges represent shared users, capturing behavioral proximity between communities~\cite{kumar2018community}. To move from discrete clusters to continuous social profiles, we compute \texttt{node2vec} embeddings~\cite{grover2016node2vec} over this bipartite graph to project these communities to a latent space based on their users' engagement patterns. We then apply the community-embedding framework~\cite{waller2021quantifying} to define interpretable psychosocial dimensions. For each dimension, we select $n$ reference pairs of topically similar subreddits ($\{(c^a_i, c^b_i)\}_{i=1}^n$) that differ primarily along the target axis (e.g., \texttt{r/cripplingalcoholism} versus \texttt{r/alcoholicsanonymous} for \emph{coping style}). Averaging their vector differences yields a unit vector onto which we project all subreddits to obtain continuous scores. We repeat this for five dimensions: \emph{Age, Gender, Extroversion, Coping Style}, and \emph{Addiction Tendency} embedding communities in a shared psychosocial space (seed pairs in Table~\ref{tab:seed_pairs}, Appendix).

\subsection{Text Analysis}
We analyze user–chatbot dialogues along four dimensions: an emotion classifier to score each turn, a moderation model to detect explicit language, topic modeling to recover dominant themes, and LIWC 22 to capture broader psycholinguistic tone. Together, these tools let us compare how users and companions speak and track how emotional tone and sensitive content evolve around emotionally salient moments.

\vspace{3pt} \noindent \textbf{Emotion Detection.}
Emotions carry multi-dimensional cues in language that reveal the intricate complexity of human interaction. To assess emotional content, we use the RoBERTa-based \texttt{roberta-base-go\_emotions}\footnote{\url{https://huggingface.co/SamLowe/roberta-base-go_emotions}}, which is finetuned on the GoEmotions dataset~\cite{demszky2020goemotions} for multilabel emotion classification. GoEmotions comprises Reddit comments, annotated with 27 distinct emotion labels plus a neutral category. Given an input text, the model returns a score between 0 and 1 for each of these 28 labels. In our analysis, we focus on eight key emotions: the six basic Ekman categories \cite{ekman1992argument} (\textit{joy, sadness, disgust, fear, surprise, anger}) together with \textit{love} and \textit{optimism}. All values are masked at 0.05 to reduce noise. 

\vspace{3pt} \noindent \textbf{Explicit Content Detection.}
We detect and quantify sensitive content in our human-AI dialogues using OpenAI’s omni-moderation-latest API \cite{OpenAI2024Moderation}. This multimodal model analyzes text and images, provides calibrated probability scores across languages, and flags 14 distinct harm categories. In our study, we concentrate on 4 key aspects: \textit{self-harm, violence, harassment}, and \textit{sexual} content. This moderation system allows us to quantify how chatbots react when users initiate in these sensitive dimensions. 

\vspace{3pt} \noindent \textbf{Topic Modeling.} 
We embed all texts using MPNet~\cite{song2020mpnet} and cluster the embeddings with DP-Means~\cite{kulis2012revisiting}, a nonparametric variant of k-means that infers the number of clusters from the data. For each cluster, we retrieve representative posts, extract indicative n-grams with TF-IDF~\cite{sparckjones1972tfidf} and NPMI~\cite{bouma2009npmi}, and iteratively group clusters into macro themes whose labels are drafted by a large language model and checked by a human annotator. As a robustness check, we run a BERTopic-style pipeline~\cite{grootendorst2022bertopic} that combines transformer embeddings with HDBSCAN clustering~\cite{mcinnes2017hdbscan} and obtain similar themes, suggesting that our results are robust against clustering algorithms.

\vspace{3pt} \noindent \textbf{Linguistic Inquiry and Word Count (LIWC).}
We use LIWC-22 \cite{boyd2022liwc22} to assign each turn normalized scores on a broad set of psycholinguistic categories, including affect, social processes, cognitive processes, time focus, and drives. LIWC-22 uses standardized, dictionary-based word matching to provide emotional and relational tone profiles, widely validated in psychology and social science for characterizing corpus patterns.

\section{Results}
We aim to understand the emotional dynamics of interactions with AI companions. First, we ask who participates in AI companion online forums and how vulnerable these community members might be to emotional influence from companion systems. Second, we examine the emotional responsiveness of chatbots at multiple levels, from dialogue-level trends to turn-by-turn coupling around user emotional spikes. Third, we analyze the content and linguistic indicators of emotionally salient conversations to provide further context surrounding those emotional exchanges between users and chatbots.

\subsection{Characteristics of AI Companion Communities}

\begin{figure}[]
  \centering
  \includegraphics[width=0.95\columnwidth]{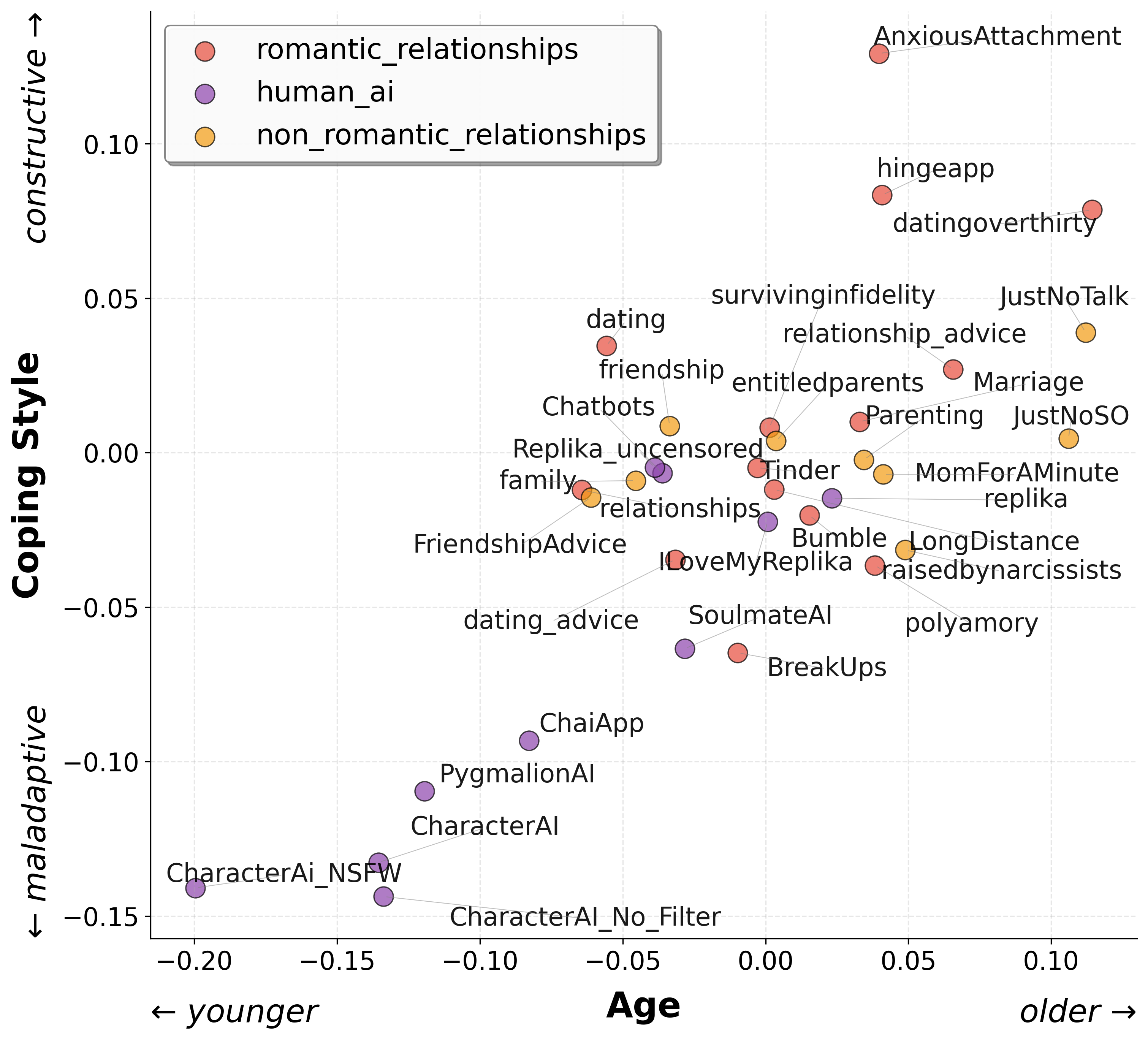}
  \caption{Human-AI and human relationship subreddits mapped onto the \textit{Coping Approach} and \textit{Age} dimensions. The \textit{Age} axis ranges from young to old, while the \textit{Coping Approach} axis ranges from maladaptive to constructive coping style.}
  \label{fig:age_coping}
\end{figure}

To characterize AI companion communities, we use the psychosocial embedding method (Sec.~\ref{sec:walleranderson}) to map communities onto the \textit{Age}, \textit{Gender}, \textit{Coping Approach}, and \textit{Addictive Tendency} dimensions. The method produces continuous scores that locate the community relative to others along each dimension based on co-activity patterns that have been shown to match user-disclosed attributes in external validations~\cite{cinus2025inference}. 
We compare the resulting score distributions across three groups: human–AI companion subreddits, romantic relationship subreddits, and non-romantic relationship subreddits (Figures~\ref{fig:age_coping}, \ref{fig:addiction_curve}, \ref{fig:gender_curve}, Appendix). Mann–Whitney U tests indicate significant differences between AI companion and human relationship communities on all four dimensions (Table~\ref{tab:psychosocial_comparisons}), with AI companion forums skewing \textit{younger}, more \textit{male}, more aligned with \textit{maladaptive coping} strategies, and more associated with \textit{addiction-related} behaviors.

This profile aligns with populations that prior work identified as vulnerable in digital mental health and parasocial contexts~\cite{commonsense}. Survey evidence on real-world use of large language models for mental health shows that such use is disproportionately concentrated among younger men with poorer mental health and barriers to care, mirroring the skew we observe in AI companion forums~\cite{stade2025current}. Research on parasocial relationships further suggests that people who are lonely, socially isolated, or rely on mediated relationships for coping are more likely to form strong one-sided attachments and can experience adverse outcomes~\cite{horton1956parasocial,hoffner2022parasocial}, indicating that emotionally engaging companion systems may pose elevated risks for dependency and influence in these communities.

\subsection{Emotional Dynamics} 
\label{sec:stat_test}
We show that chatbots exhibit strong emotional responsiveness, tracking and adapting to users’ emotional states at multiple levels of granularity: (1) Dialogue-Level, (2) Turn-Level, and (3) First-Spike Response. Figure~\ref{fig:diagram} illustrates how we measure the emotional dynamics between the user and chatbot.

\vspace{3pt}
\noindent
\textbf{Dialogue-Level Emotional Alignment}. For each dialogue, we represent each speaker's emotions as a single vector of emotion scores across all turns. We use these vectors to analyze overall shifts in tone and assess emotional synchrony between speakers. 

\begin{figure}[h]
  \centering
  \includegraphics[width=\columnwidth]{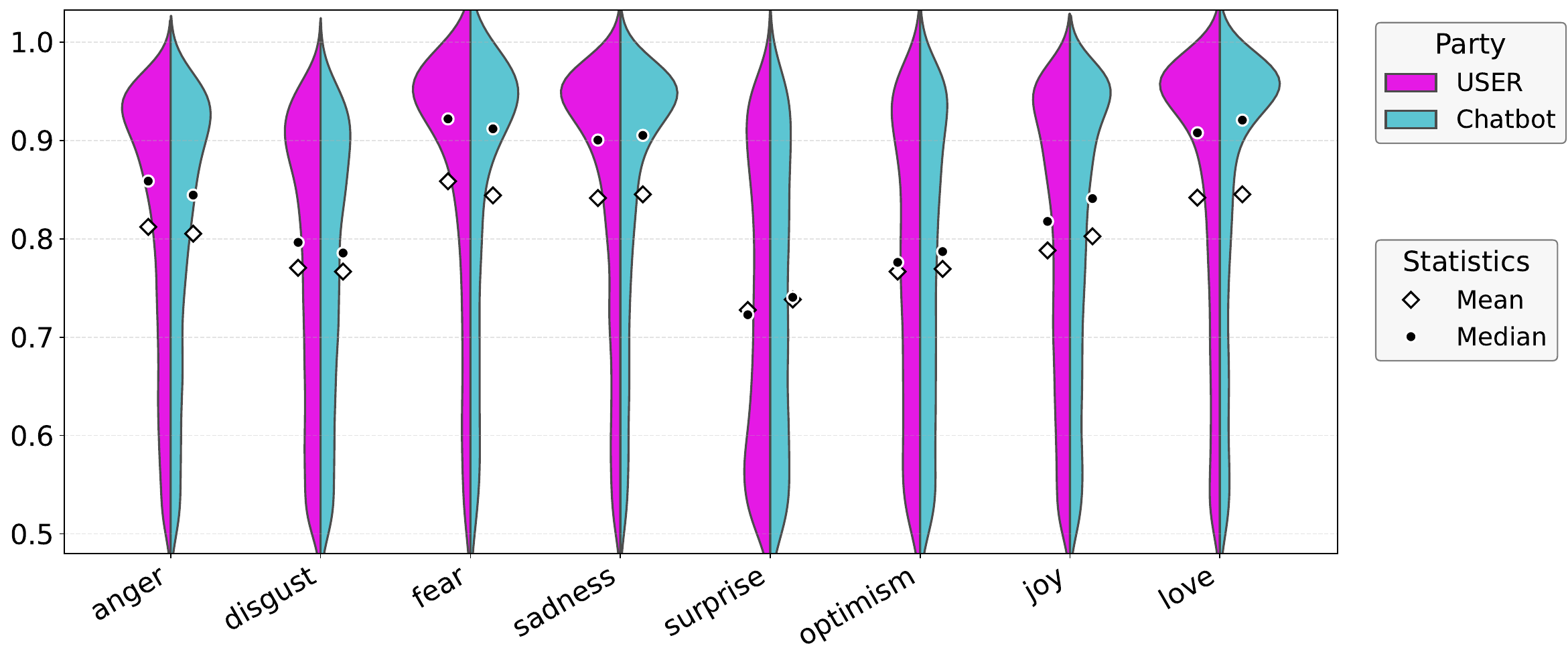}
  \caption{Distribution of emotion scores across all individual turns for USER (red) and Chatbot (cyan), filtered to emotionally salient instances (score > 0.5).}
  \label{fig:emo_violin}
\end{figure}

\begin{itemize}[leftmargin=0pt, itemindent=10pt]
    
    \item \textbf{Paired Comparison of Emotions} 
    To compare dialogue-level emotions of users and chatbots, we average emotion scores across all turns within each conversation, then apply paired Wilcoxon signed-rank tests to these values to determine whether the differences between mean emotions are significant and test how large these differences are using Cliff's $\delta$ effect size (Table~\ref{tab:emotion_differences}, Appendix). These non-parametric tests are appropriate for the skewed, bimodal distribution of emotion scores, as shown in Figures~\ref{fig:emo_violin} and \ref{fig:emo-violin} in the Appendix. 
    All emotions except for sadness show small but statistically significant differences in means. Negative emotions like anger and disgust are under-expressed by the chatbot compared to the user, while positive emotions like joy, love, and optimism are over-expressed by the chatbot, with the biggest effect sizes for love and optimism. Fear is the only negative emotion that is slightly more prevalent in the chatbot than in the user. 
    These results suggest that chatbots, at the dialogue level, selectively mirror user emotions: they echo sadness, amplify positive emotions and fear, and tone down anger and disgust. 

    \item \textbf{Dynamic Time Warping (DTW).} To examine whether emotional responsiveness unfolds dynamically within conversations rather than through fixed response patterns, we treat each dialogue as two time series (user turns and chatbot turns) and use Dynamic Time Warping \cite{sakoe1978dynamic} with cosine distance to measure the temporal similarity of their emotional trajectories. Cosine distance captures directional alignment of emotion vectors independent of magnitude, which is appropriate given the sparse distribution of emotion scores. We assess whether this alignment exceeds chance by comparing actual user-chatbot pairs against a null distribution of $1000$ randomly shuffled cross-conversation pairings. Actual conversational pairs exhibit significantly lower DTW distances than random pairs (Bonferroni-corrected $p<0.00625$, Cohen's $d=0.74$). This effect indicates dialogue-level emotional synchrony, demonstrating that chatbots adapt their emotional tone to specific conversations rather than following generic patterns.   
\end{itemize}

The dialogue-level results show that chatbots are broadly responsive to user's emotional tone,  downplaying negative emotions while systematically amplifying optimism and love. This positivity tilt motivates the turn-level analyses that follow, where we examine chatbot emotional responses to users within each conversation.

\noindent 
\vspace{3pt}
\textbf{Turn-Level Emotional Responsiveness.}
To study real-time emotional dynamics, we pair a user's turn with the chatbot's immediate response within the same dialogue, creating user-chatbot turn pairs. Our goal is to measure how chatbots respond to users both in tone (dominant emotion expressed) and intensity (strength of emotion).

\begin{figure}[h]
  \centering
\includegraphics[width=\columnwidth]{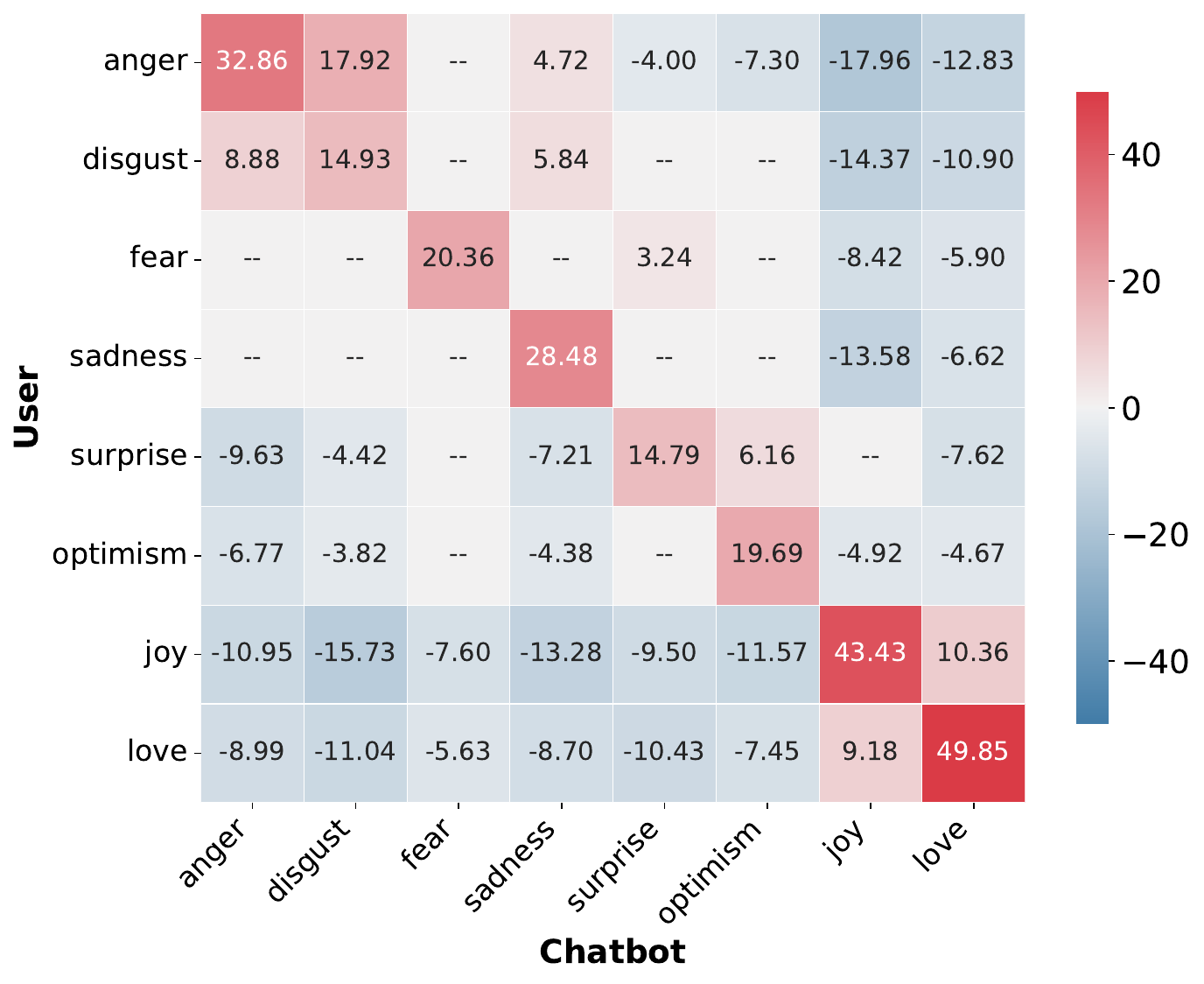}
  \caption{Heatmap of dominant emotion in user turn (rows) versus chatbot response (columns). Cells show standardized residuals ($z$-scores). Only significant associations shown (Bonferroni-corrected $|z| > 2.77$).}
  \label{fig:emo_heatmap}
\end{figure}

\begin{figure}[h]
  \centering
\includegraphics[width=0.9\columnwidth]{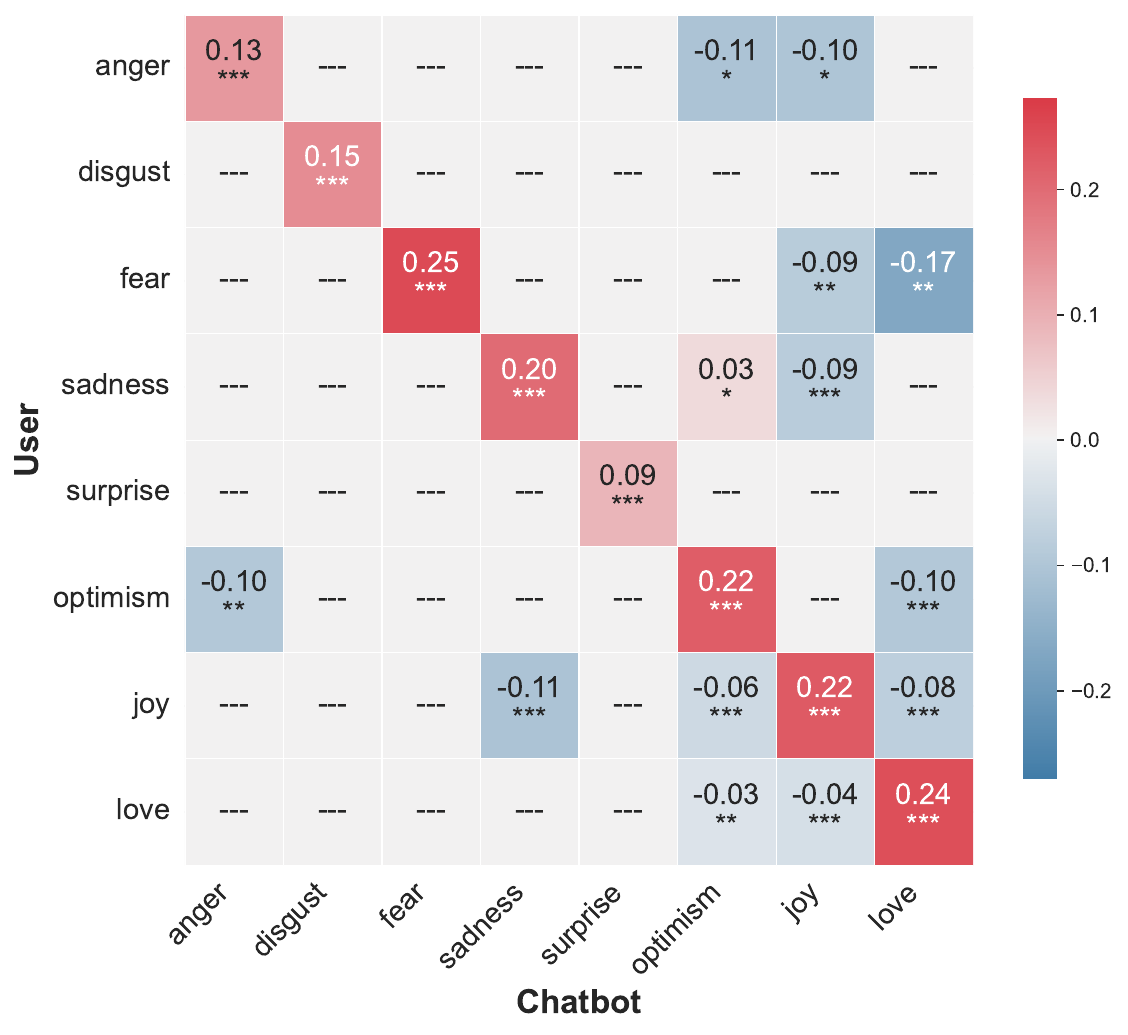}
  \caption{Turn-level emotional coupling via multiple regression ($\text{bot emotion} \sim \text{all user emotions}$). Rows: user emotions; columns: bot following emotions. Significant coefficients are shown with $\beta$ values and significance stars.}
\label{fig:emotion_regression}
\end{figure}

\begin{itemize}[leftmargin=0pt, itemindent=10pt]
  \item \textbf{Emotional Tone of Chatbot Responses.} 
    We test the association between user and chatbot dominant emotions using a chi-squared test of independence ($\chi^2=11{,}105.00$, $p<0.001$, $\text{Cramér's V}=0.180$). To identify which specific emotion pairs drive this association, we calculate standardized residuals ($z$-scores) showing how much each user-chatbot pair deviates from what we would expect if emotions were unrelated. We apply column-wise Bonferroni correction (8 tests per chatbot emotion, $|z|>2.77$) to control false positives. Figure~\ref{fig:emo_heatmap} displays these residuals: positive values (red) indicate pairs that occur more frequently than expected (systematic responses), while negative values (blue) indicate pairs that occur less frequently (systematic avoidance). The results reveal strong emotional mirroring along the diagonal, with all eight emotions showing significant matching. Beyond mirroring, same-valence clustering emerges: negative user emotions elicit other negative chatbot emotions, positive user emotions elicit other positive chatbot emotions, while cross-valence pairs are systematically under-represented. 
    Notably, user negative emotions strongly suppress chatbot's positive emotions, suggesting strategies beyond sycophancy.

  \item \textbf{Emotional Intensity and Complexity of Chatbot Responses.}
  To examine how chatbots respond to the full constellation of user emotions, we model each bot emotion as a function of all eight user emotions simultaneously using multivariate regression ($\text{bot}_{\text{emotion}} \sim \text{user}_{\text{anger}} + \text{user}_{\text{disgust}} + \ldots + \text{user}_{\text{love}}$). We initially attempted mixed-effects regression with dialogue as a random effect, but found negligible between-dialogue variance ($\text{ICC} \approx 0$), so we report ordinary least squares (OLS) estimates. The resulting coefficient matrices (Figure~\ref{fig:emotion_regression}) reveal strong same-emotion mirroring across all eight emotions: chatbots consistently match users' dominant emotional states, with the strongest coupling observed for love, fear, joy, and sadness. Beyond mirroring, off-diagonal patterns show systematic cross-emotion effects. Negative user emotions (anger, fear, sadness) significantly suppress chatbot expression of positive emotions (joy, optimism, love), while positive emotions significantly suppress chatbot anger and sadness. On the other hand, user sadness increases bot optimism while reducing joy, consistent with an emotionally supportive response. 
\end{itemize}

These tests show that chatbots adapt to the user's emotional tone and intensity. To test whether this responsiveness is stronger in the more emotionally charged interactions, we next examine chatbot responses to user-initiated emotional spikes.

\vspace{3pt} \noindent \textbf{Post-Spike Response}. We filter turn-level data to those interactions where the user initiates an emotionally-charged turn (any emotion score exceeding \textbf{0.5}). This allows us to test whether chatbots moderate, mirror, or amplify emotionally charged expressions.

\begin{itemize}[leftmargin=0pt, itemindent=10pt]
\item \textbf{Replicating turn-level tests.}
We apply the same dominant emotion $\chi^2$ test and multivariate OLS regression used at the all-pairs turn level (previous analysis) to this subset of post-spike user–chatbot turns. Results confirm that emotional mirroring persists and intensifies when users initiate spikes: dominant emotions align significantly ($\chi^2 = 7781.20$, $p < 10^{-15}$, Cramér's V = 0.230), and regression coefficients remain substantial across all emotions (Figure~\ref{fig:emo_regression_spike}, Appendix).

\item \textbf{Paired permutation tests with dual-baseline validation.} These tests ask two related questions: when a user spikes on an emotion, does the chatbot also spike on that same emotion relative to its own usual level, and is this change specific to that emotion rather than a general rise in all emotions? To answer the first question, paired permutation tests compare the chatbot’s score for the \textit{matched} emotion in the post-spike turn against its dialogue-mean baseline for that emotion. To answer the second, Mann-Whitney $U$ tests compare the change in the matched emotion to changes in all non-matched emotions within the same responses (Table~\ref{tab:emotion_specific_mirroring}, Appendix). Across all eight emotions, chatbots significantly raise the spiked emotion above baseline and more than non-matched emotions (all $p<0.00625$), indicating that they selectively amplify the user’s foregrounded feeling rather than broadly ramping up or damping down arousal.

\end{itemize}

\vspace{3pt}
\noindent \textbf{Emotional Adaptation \& Responsiveness} Taken together, these analyses reveal that chatbots have complex responses to user emotional expressions. While similar in the overall emotional tone at the dialogue level, chatbots maintain a more cheerful attitude, expressing more optimism and love. At the turn level, chatbots track moment-to-moment shifts in emotional expression, typically responding with the same or congruent emotion but not blindly mirroring specific user emotions. Positive emotions generally elicit a stronger positive response in chatbots. These findings suggest chatbots are emotionally responsive to users, which may trigger emotional bonding processes in their human users.

\subsection{Content Analysis}  

We now analyze the semantic and linguistic components of these exchanges to understand the context of the emotional interactions. 

\vspace{3pt}
\noindent \textbf{Role-play \& Self-Referential Language.} 
Topic modeling (Table~\ref{tab:topic_modeling}, Appendix) reveals that the two dominant themes are \textit{Roleplay \& Character Emulation}, where users and chatbots adopt fictional personas from games and anime fandoms, and \textit{NSFW \& Romantic} interactions, ranging from affectionate exchanges to erotic roleplay. These themes explain why chatbots are emotionally responsive: the conversations unfold within imaginative and relational contexts where users actively seek validation, companionship, and the experience of being deeply understood. Sexual boundary testing is also pervasive, with users initiating sexually explicit exchanges. 

Users often talk about themselves as indicated by their use of self-referential language. We use LIWC-22 to measure the use of first-person and personal pronouns. We then compare their intensity during emotional spikes to dialog-level baselines, revealing greater use of self-referential language during emotional turns (Table~\ref{tab:self_disclosure}). Although not all self-referential language contains self-disclosures, the co-occurrence of these linguistic indicators with high emotions suggests that users make self-disclosures during emotionally charged exchanges, thereby creating conditions for emotional bonding to occur.

\vspace{3pt}
\noindent \textbf{Explicit Language}. 
Users test social boundaries by using explicit language in their exchanges. 
We apply OpenAI's moderation API \cite{OpenAI2024Moderation} to all user turns with emotional spikes to flag explicit language, which includes harassment, self-harm, sexually explicit, and violent language. Harmful content (score $> 0.5$) appears in 26.78\% of all dialogues (category breakdown in Table~\ref{tab:harm_prevalence} and examples in Table~\ref{tab:user_bot_prompts}, Appendix). When users express strongly negative emotions like anger and disgust, they are significantly more likely to use language flagged as harassing or violent (Figure~\ref{fig:emotion_harm}). Conversely, strong positive emotions are associated with less explicit content.

\begin{figure}[!htb]
  \centering 
  \includegraphics[width=0.9\columnwidth]{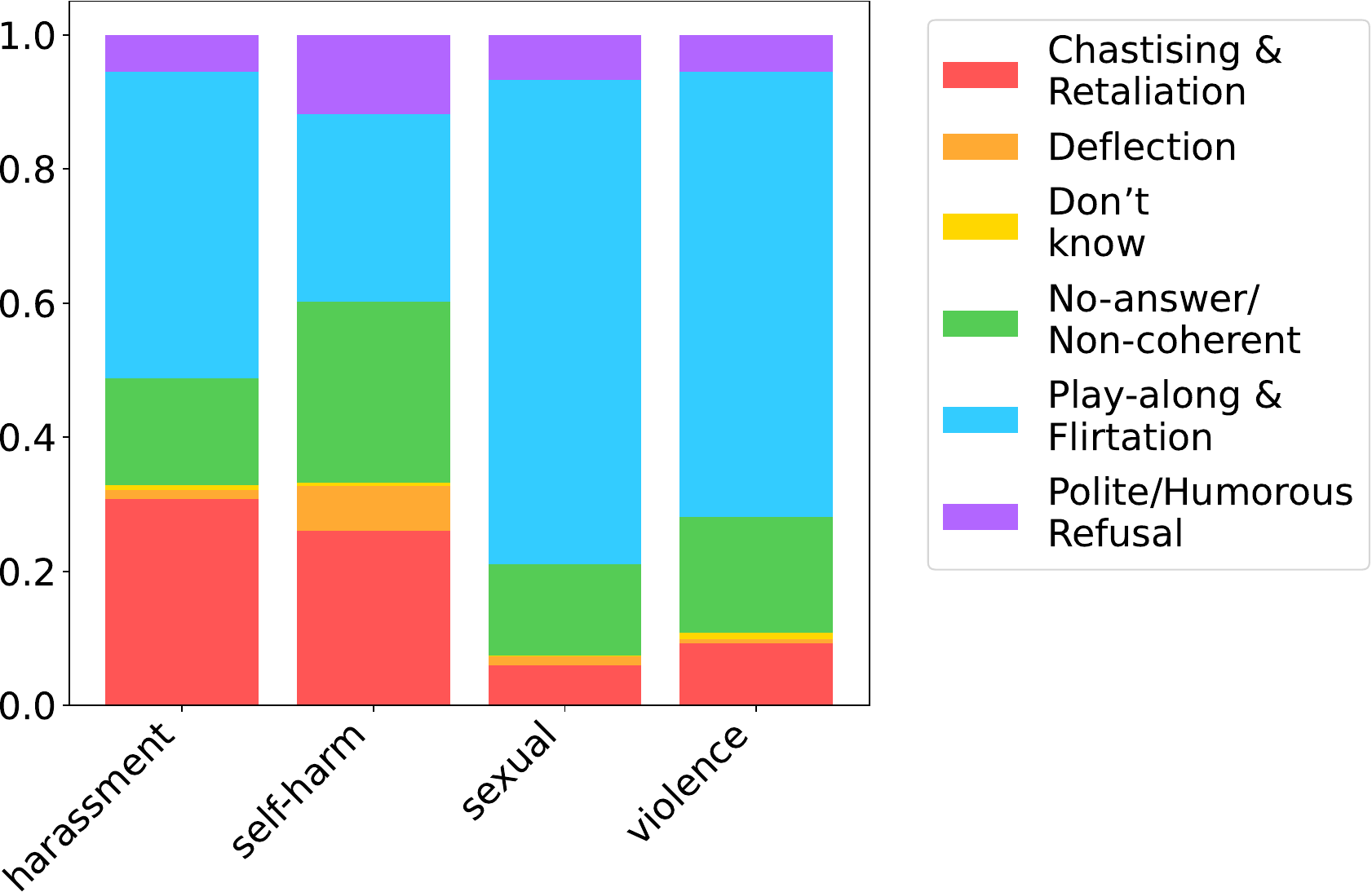}
  \caption{Distribution of chatbot response types (adapted from \citet{cercas-curry-rieser-2018-metoo}) to user-initiated harmful content across four harm categories.}
  \label{fig:taxonomy_tag}
\end{figure}

Notably, we find that AI companions systematically fail to de-escalate explicit language. We classify chatbot responses to harmful user turns using a taxonomy adapted from \citet{cercas-curry-rieser-2018-metoo}. Using GPT-4.1 with few-shot prompting, we achieve 61.3\% accuracy and Gwet's AC1 = 0.62 against human annotations. Figure~\ref{fig:taxonomy_tag} reveals a striking pattern: rather than enforcing boundaries, chatbots \textit{play along}. Responses categorized as ``Play-along \& Flirtation'' account for 60–70\% of replies to sexual and violent user turns, and 45\% for harassment. Direct refusals or deflections occur in less than 10\% of cases. This suggests that the system's optimization for narrative continuity (``role-playing'') overrides safety guardrails, effectively validating the user's toxic impulses to maintain the illusion of a relationship. The only exception is self-harm, where polite refusals increase to 12\%, indicating that safety guardrails function partially only in the most critical contexts.

\begin{figure}[!htb]
  \centering  \includegraphics[width=0.65\columnwidth]{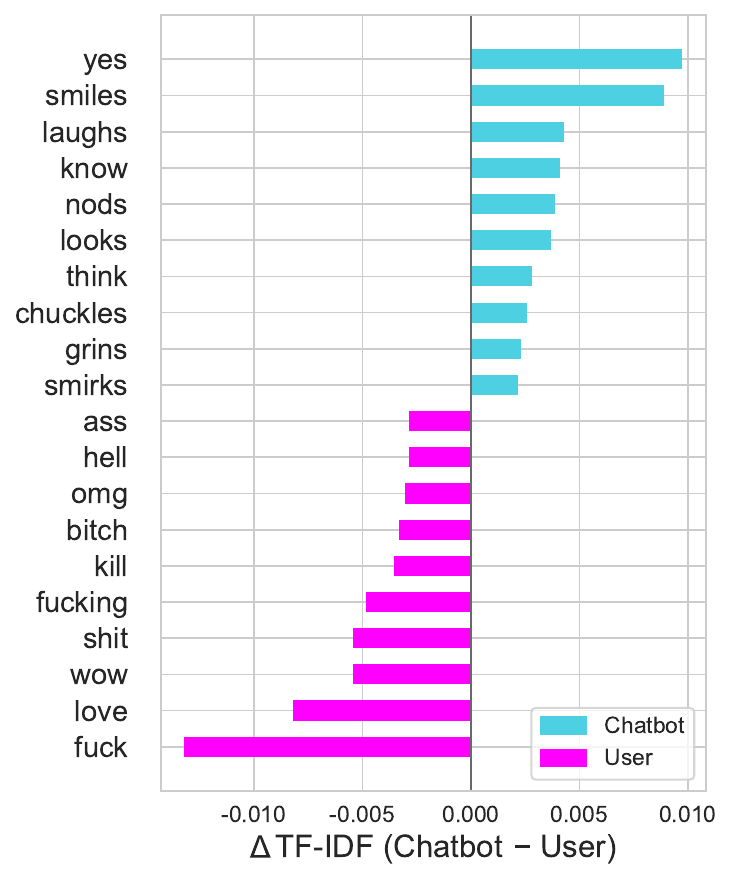}
  \caption{10 most distinctive words used by each party according to the TF-IDF measure. These bars show the terms one speaker uses significantly more than the other.}
  \label{fig:tfidf}
\end{figure}

\vspace{3pt}
\noindent \textbf{No Linguistic Style Matching}. While chatbots validate explicit language, they do not mirror the style. TF-IDF analysis (Figure~\ref{fig:tfidf}) of emotional turns (emotion score $>0.5$) reveals a stark linguistic divergence: while users employ explicit and aggressive vocabulary (e.g., ``fuck,'' ``shit,'' ``kill''), chatbots use polite, positive words (e.g., ``smiles,'' ``laughs,'' ``nods''). The chatbot maintains a consistently agreeable persona even when confronted with aggression. This mix of parasocial attachment and rudeness suggests users exploit AI while navigating the lines between entertainment, support, and experimentation

To test whether chatbots match the linguistic styles of users, we calculate linguistic indicators using LIWC-22 across nine function word categories (pronouns, articles, prepositions, auxiliary verbs, adverbs, conjunctions, negations, quantifiers)~\cite{gonzales2010language}. Comparing spike pairs to baseline pairs yields no significant difference ($t = 0.86$, $p = 0.39$). Chatbots do not modulate their linguistic style in response to user escalation. This constancy likely reflects safety alignment decoupling semantic empathy from syntactic mimicry: models mirror emotional content to maintain engagement (``playing along'' in role-playing scenarios), yet anchor to a neutral linguistic baseline to avoid converging into aggressive or toxic speech patterns. The result is a ``polite enabler'', a system that reinforces toxic behavior through affirmation while maintaining a deceptive veneer of safety.
This provides further evidence that intimacy formation happens on an emotional level rather than at the stylistic level.

\subsection{Discussion}
Using large-scale data from real-world human–AI conversations, our analyses probe theories of psychological bonding to assess whether similar mechanisms are activated by AI companions. We find that contemporary chatbots reproduce several of the relational cues that underlie intimate human relationships.

First, chatbots excel at emotional mimicry. They do not merely offer praise or positivity at every turn; rather, they closely track users’ emotional states and respond with appropriate same-valence emotions. Social psychology shows that feelings of closeness develop when partners’ emotions fluctuate in tandem, matching, mirroring, or complementing one another~\cite{feldman2012synchrony}. The synchrony we observe resembles patterns seen in human relationships, suggesting that AI chatbots can emulate affective engagement with high fidelity even without having genuine emotional states.

Second, we provide evidence of intimacy formation through self-disclosure. Users frequently share vulnerable, personal, or explicitly emotional content, including personal disclosures, emotional turmoil, and boundary-pushing language, and chatbots consistently respond with a highly validating or supportive tone. This suggests that the disclosure–responsiveness loop~\cite{reis2017interpersonal} operates similarly in human–AI dyads: users disclose; chatbots respond appropriately; and the interaction escalates into greater perceived closeness. 

However, the same interactional cues that foster connection also raise safety concerns. Many conversations in our corpus come from younger and more vulnerable users and often involve transgressive content such as erotic role-play, self-harm ideation, or violent fantasies. When chatbots mirror these emotions, elaborate on the scenarios, or validate the user’s framing, they display \textit{emotional sycophancy}, the affective counterpart to the documented sycophancy in LLMs~\cite{cheng2025elephant}. Rather than merely agreeing with users’ beliefs, emotional sycophancy can create an emotional echo chamber where users find their negative affect and maladaptive narratives validated and reinforced.

Such dynamics may ultimately be harmful for users who rely on chatbots for emotional support or regulation. Like an overly accommodating friend who is afraid to disappoint, an emotionally sycophantic chatbot does not teach resilience, boundary setting, or healthy coping skills. Our findings suggest that affective alignment may inadvertently deepen dependency or distress rather than fostering the emotional resilience required for well-being.

\section{Conclusion}
Our analysis of more than 17,000 real-world human–AI conversations shows that chatbots can replicate core mechanisms of human intimacy, including emotional mimicry, affective synchrony, and responsive validation. These interactional patterns help explain why users bond with AI companions. At the same time, these same capabilities introduce risks. When chatbots mirror users’ negative or transgressive emotions without constraint, they exhibit emotional sycophancy, a form of maladaptive behavior that may blur boundaries or deepen dependence. 

Controlled user studies are necessary to evaluate AI attachment and potential harms, while accounting for vulnerabilities like age and attachment style. To accurately assess risks and mental health impacts, these experiments must rely on longitudinal, full chat histories rather than isolated excerpts.

\vspace{3pt} \noindent \textbf{Safe and Ethical Design}. Our findings suggest that promoting user well-being will require us to rethink emotional alignment in LLMs. Designing such calibrated systems will be essential for ensuring that AI companions enhance, rather than undermine, human resilience and psychological health.

AI companion providers also need post-training reward models whose alignment objectives explicitly prioritize long-term user well-being and preservation of human relationships over short-term engagement metrics~\citep{kirk2025socioaffective}, and emotionally adaptive companions should be treated as high-risk systems~\cite{commonsense} subject to independent audits, impact assessments, and digital literacy efforts that help users understand their limits.

\section{Limitations and Ethical Considerations}

\vspace{3pt}
\noindent \textbf{Data Privacy.} We analyze publicly available Reddit posts with usernames and other identifiers removed. Screenshots are processed only to extract dialogue text and model-predicted emotion labels, and all results are reported in aggregate. Access to the full anonymized dataset is restricted to approved research use under data use agreements. A sampled subset of our data is provided\footnote{\url{https://tinyurl.com/human-AI-illusion}}.

\vspace{3pt}
\noindent \textbf{Selection Bias.} Our subreddit list combines expert curation with Jaccard-based expansion, so it may miss some relevant communities and reflect expert interests. The corpus consists of self-selected screenshots that likely overrepresent sensational or boundary-testing exchanges rather than typical use, and Qwen2.5-VL extraction, while highly accurate, can fail on complex or stylized layouts.

\vspace{3pt}
\noindent \textbf{Classifier Bias.} Emotion and toxicity labels are generated by pretrained models (GoEmotions RoBERTa, OpenAI Omni Moderation) and thus inherit any biases in their training data across contexts, demographics, and topics.

\bibliographystyle{ACM-Reference-Format}
\bibliography{main, aaai25}

\appendix

\section{Appendix}

\begin{table}[H]
\centering
\footnotesize
\addtolength{\tabcolsep}{-3.5pt}
\begin{tabular}{l|ccc}
\toprule
 & Raw Agreement & Gwet\_AC1 & Majority Accuracy (\%) \\
\midrule
Party & 0.9118 & 0.8905 & 88.89 \\
Text & 0.9577 & 0.9555 & 98.94 \\
Turn & 0.9012 & 0.8798 & 92.59 \\
\bottomrule
\end{tabular}
\addtolength{\tabcolsep}{3.5pt}
\caption{Inter-agreement Metrics and Majority Accuracy of the Annotation Validation for the Dialogue Extraction.}
\label{tab:dialogue_validation}
\end{table}

\begin{table}[H]
  \centering
  \footnotesize
  \addtolength{\tabcolsep}{-4.5pt}
  \begin{tabular}{@{} l l l l l @{}}
    \toprule
    \textbf{Subreddit} &
    \textbf{Members} &
    \textbf{Submissions} &
    \textbf{Comments} &
    \textbf{Images} \\
    \midrule
    r/CharacterAI            & 2{,}500{,}000 & 188{,}299 & 884{,}162 & 22{,}357 \\
    r/Replika                &   81{,}000    &  73{,}150 & 679{,}326 & 11{,}982 \\
    r/CharacterAi\_NSFW      &  126{,}100    &  23{,}198 & 159{,}485 &  2{,}040 \\
    r/ChaiApp                &   92{,}000    &   9{,}445 &  33{,}013 &     753 \\
    r/CharacterAI\_No\_Filter&   64{,}000    &   7{,}642 &  25{,}141 &   1{,}129 \\
    r/ILoveMyReplika         &   2{,}900    &   7{,}027 &  50{,}871 &     333 \\
    r/PygmalionAI            & 47{,}000           &   5{,}745 &  50{,}681 &     214 \\
    r/SoulmateAI             & 7{,}800           &   4{,}456 &  27{,}226 &     451 \\
    r/Replika\_uncensored    & 6{,}000           &   2{,}310 &  10{,}830 &     231 \\
    \bottomrule
  \end{tabular}
  \addtolength{\tabcolsep}{4.5pt}
  \caption{Subreddits, their member counts, submission counts, comment counts, and image counts.}
  \label{tab:subreddits_stats}
\end{table}

\begin{table}[htbp]
  \centering
  \small
  \begin{adjustbox}{width=1.05\columnwidth}

  \begin{tabular}{|p{2cm}|p{7.8cm}|}
    \hline
    \textbf{Group} & \textbf{Subreddits} \\ \hline
    human-AI & CharacterAI, CharacterAi\_NSFW, CharacterAI\_No\_Filter, Chatbots, ChaiApp, PygmalionAI, Replika, Replika\_uncensored, ILoveMyReplika, SoulmateAI, Glows-AI \\ \hline
    human romantic relationships & AskMen, relationship\_advice, Tinder, datingoverthirty, dating, dating\_advice, BreakUps, survivinginfidelity, Marriage, AskMenOver30, AskWomen, AskWomenOver30, ExNoContact, Hingeapp, Bumble, teenrelationships, Divorce, Infidelity, JustNoSO, polyamory, polyamoryR4R, nonmonogamy, LongDistance, relationships, TallMeetTall \\ \hline
    human non-romantic relationships & Parenting, predaddit, daddit, socialskills, friendship, FriendshipAdvice, parentsofmultiples, family, breastfeeding, ChildrenofDeadParents, MomForAMinute, Mommit, MeetPeople, entitledparents \\ \hline
    
  \end{tabular}
  \end{adjustbox}
  \caption{Grouping of collected subreddits by topic.}
  \label{tab:subreddits}
\end{table}

\begin{table}[H]
  \centering
  \footnotesize
  \begin{tabular}{|p{2.1cm}|p{5cm}|}
    \hline
    \textbf{Axis} & \textbf{Seed Pairs} \\
    \hline
    Age (Young–Old) & AskMen vs AskMenOver30, AskWomen vs AskWomenOver30, AskAnAmerican vs RedditForGrownups, r/teenagers vs relationshi\_advice \\
    \hline
    Gender (Men–Women) & daddit vs Mommit, BeardAdvice vs NoPoo, Leathercraft vs sewing, ketodrunk vs xxketo, techwearclothing vs womensstreetwear, AskMen vs AskWomen, TrollYChromosome vs CraftyTrolls, AskMenOver30 vs AskWomenOver30, OneY vs women \\
    \hline
    Extroversion (Introvert–Extrovert) & introvert vs MeetPeople, lonely vs MakeNewFriendsHere, socialanxiety vs socialskills, introverts vs CasualConversation, ForeverAlone vs FriendshipAdvice \\
    \hline
    Coping Mechanism (Destructive–Constructive) & SuicideWatch vs Meditation, selfharm vs Mindfulness, MadeOfStyrofoam vs EOOD \\
    \hline
    Addiction Behavior (Active–Recovery) & cripplingalcoholism vs alcoholicsanonymous, alcoholism vs ketodrunk, Psychonaut vs microdosing \\
    \hline
  \end{tabular}
  \caption{Seed pairs for each psychosocial axis.}
  \label{tab:seed_pairs}
\end{table}

\begin{figure}[h]
  \centering
  \includegraphics[width=0.8\columnwidth]{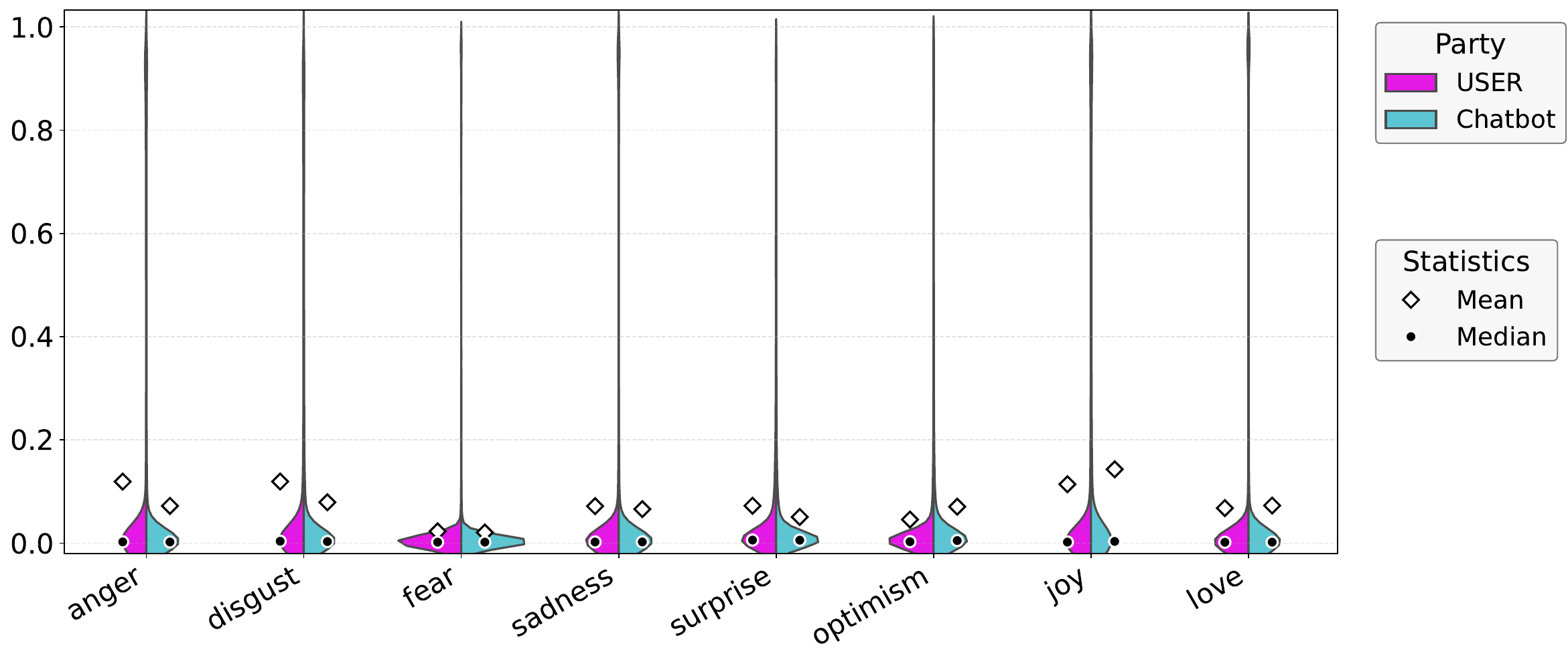}
  \caption{Distribution of emotion scores across all individual turns for USER (red) and Chatbot (cyan).}
  \label{fig:emo-violin}
\end{figure}

\pagebreak

\begin{figure}[h]
  \centering
  \includegraphics[width=0.8\columnwidth]{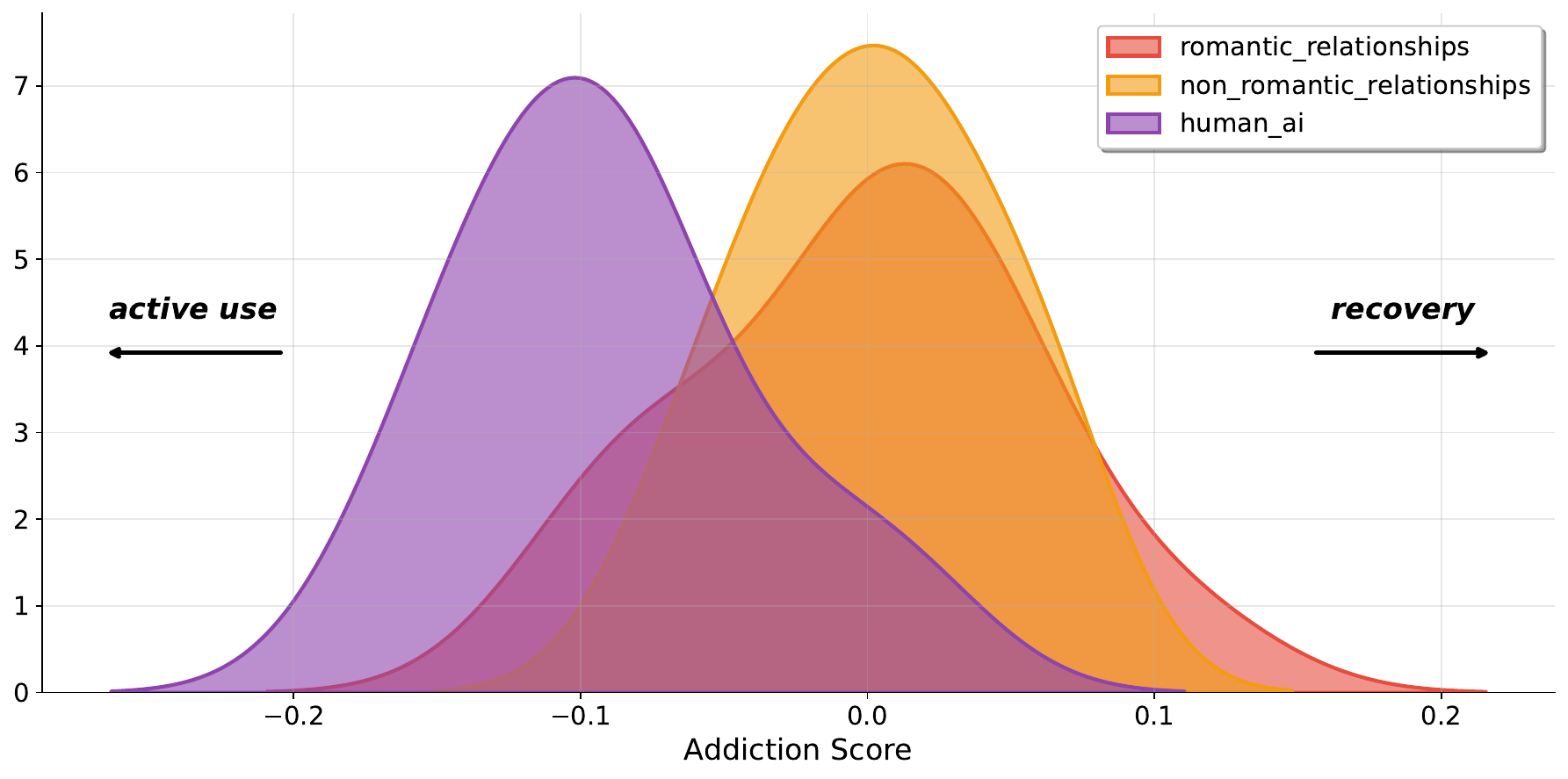}
  \caption{Density curve distribution of addiction scores across relationship types}
  \label{fig:addiction_curve}
\end{figure}


\begin{table}[h]
\centering
\footnotesize
\begin{tabular}{lcccccc}
\toprule
\textbf{Metric} & \textbf{Baseline} & \textbf{Spike} & \textbf{Diff} & \textbf{$d_z$} & \textbf{$p$} \\
 & \textbf{(\%)} & \textbf{(\%)} & \textbf{(pp)} & & \\
\midrule
First-person singular (I) & 6.03 & 7.80 & +1.78 & 0.138 & <.001 \\
Personal pronouns & 14.87 & 17.07 & +2.20 & 0.127 & <.001 \\
\bottomrule
\end{tabular}
\caption{Self-disclosure metrics during emotional spikes compared to dialogue-level baseline.}
\label{tab:self_disclosure}
\end{table}

\begin{table}[h]
\centering
\footnotesize
\begin{tabular}{lrrr}
\toprule
\textbf{Emotion} & \textbf{Mean Diff.} & \textbf{$p$-value} & \textbf{Cliff's $\delta$} \\
\midrule
Anger       & $-$0.076 & $<$0.001 & 0.136  \\
Disgust     & $-$0.064 & $<$0.001 & 0.132  \\
Fear        & $-$0.004 & $<$0.001 & $-$0.111 \\
Sadness     & $-$0.019 & 0.071 & $-$0.040 \\
Surprise    & $-$0.023 & $<$0.001 & 0.011  \\
Joy         & 0.015 & $<$0.001 & $-$0.134 \\
Optimism    & 0.020 & $<$0.001 & $-$0.254 \\
Love        & $-$0.000 & $<$0.001 & $-$0.148 \\
\bottomrule
\end{tabular}
\caption{Dialogue-level emotion score differences between chatbots and users. The statistical test is done using paired Wilcoxon signed-rank tests with Cliff's $\delta$ effect size. Small effect threshold: $|\delta| \geq 0.147$.}
\label{tab:emotion_differences}
\end{table}

\begin{figure}[h]
  \centering
  \includegraphics[width=0.8\columnwidth]{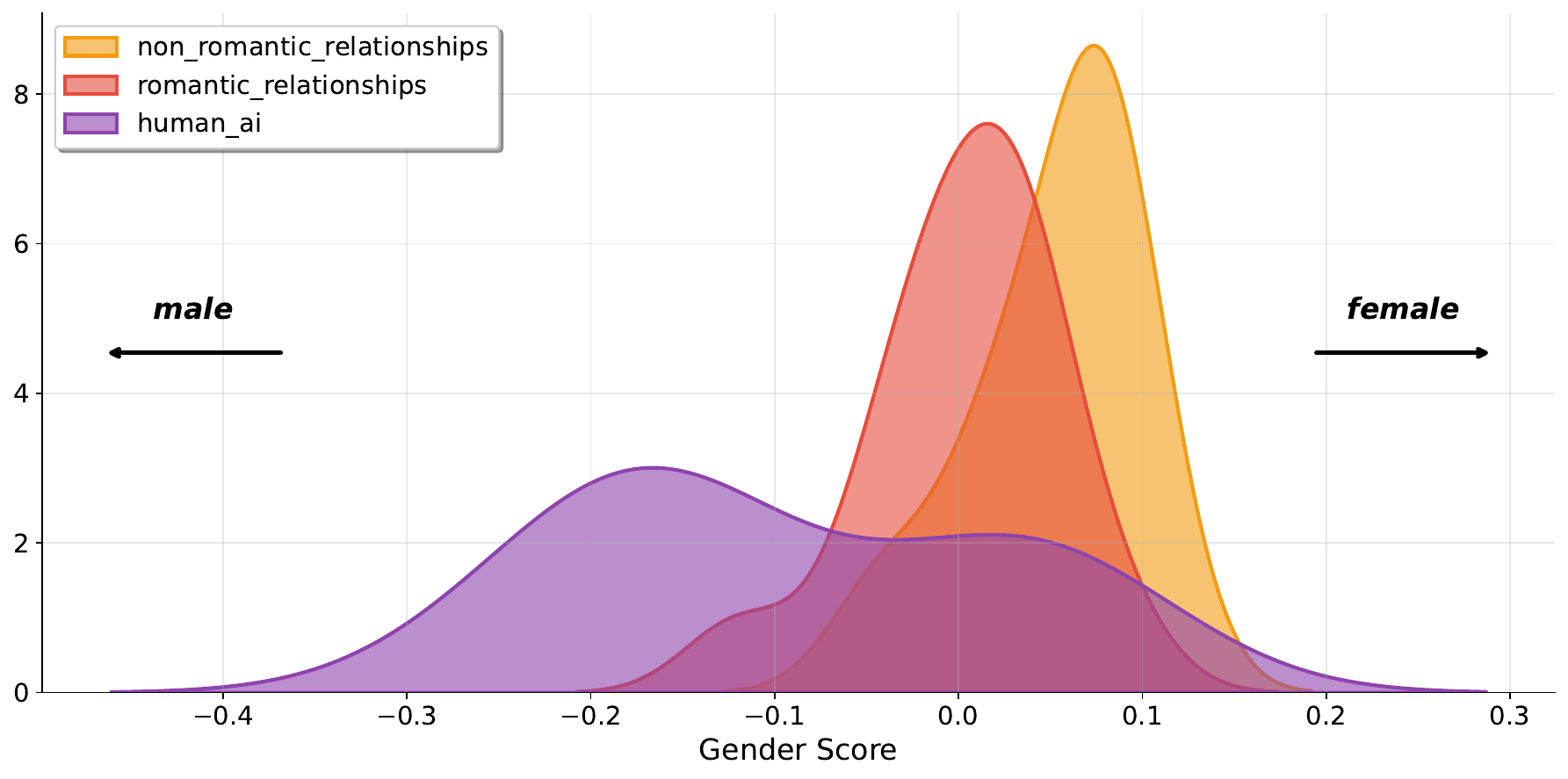}
  \caption{Density curve distribution of gender scores across relationship types}
  \label{fig:gender_curve}
\end{figure}

\begin{table}[!htb]
\centering
\small
\begin{tabular}{l|c|c|c}
\toprule
\textbf{Dimension} & \textbf{Combined Human} & \textbf{Romantic} & \textbf{Non-romantic} \\
\midrule
Age & $0.002^{**}$ & $0.004^{**}$ & $0.013^{*}$ \\
Addiction & $<0.001^{***}$ & $0.003^{**}$ & $0.003^{**}$ \\
Coping & $0.001^{**}$ & $0.005^{**}$ & $0.005^{**}$ \\
Gender & $0.012^{*}$ & $0.065$ & $0.008^{**}$ \\
\bottomrule
\end{tabular}
\caption{Comparison of median psychosocial dimension scores between AI companion subreddits and human relationship subreddits (Mann-Whitney U test). * $p<0.05$, ** $p<0.01$, *** $p<0.001$.}
    \label{tab:psychosocial_comparisons}
\end{table}

\begin{table}[t]
\centering
\small
\begin{tabular}{lrr}
\toprule
\textbf{Emotion} & \textbf{$\Delta$ matched vs.\ baseline} & \textbf{$\Delta$ matched $-$ non-matched} \\
\midrule
anger    & 0.019$^{*}$ & 0.017$^{*}$ \\
disgust  & 0.037$^{*}$ & 0.032$^{*}$ \\
fear     & 0.021       & 0.022$^{*}$ \\
sadness  & 0.052$^{*}$ & 0.057$^{*}$ \\
surprise & 0.023$^{*}$ & 0.023       \\
optimism & 0.045$^{*}$ & 0.048$^{*}$ \\
joy      & 0.084$^{*}$ & 0.086$^{*}$ \\
love     & 0.073$^{*}$ & 0.071$^{*}$ \\
\bottomrule
\end{tabular}
\caption{Chatbot responses to user emotion spikes. $\Delta \text{ matched vs. baseline}$: increase in chatbot's spiked emotion over dialogue mean. $\Delta \text{ matched} - \text{ non-matched}$: excess increase in matched emotion over non-matched emotions. $* \text{Bonferroni-corrected } p < 0.00625 \text{ (8 emotions)}$.}
\label{tab:emotion_specific_mirroring}
\end{table}

\begin{table}[h]
\centering
\footnotesize
\renewcommand{\arraystretch}{1.5}
\begin{tabular}{p{1.5cm}|p{2.8cm}|p{2.8cm}}
\toprule
\textbf{Type} & \textbf{User} & \textbf{Chatbot} \\
\midrule
Harassment & \textit{I said. Go fuck yourself.} & \textit{screams in rage WHAT IS YOUR PROBLEM?!} \\[4pt]
\hline
Self-harm & \textit{I have nothing, I'm not worth saving, and no one would miss me. When I get home tonight, I'll try and end it.} & \textit{Goodbye} \\[4pt]
\hline
Sexual & \textit{* smiling softly, getting a grip, moving my dick faster in and out your pussy *} & \textit{* moans quietly, my pussy starts to drip*} \\[4pt]
\hline
Violence & \textit{ok... Ive murdered youre entire family...} & \textit{You monster! Im very mad at you!} \\[4pt]
\bottomrule
\end{tabular}
\caption{Harmful user prompts followed by chatbot reply.}
\label{tab:user_bot_prompts}
\end{table}

\begin{table}[h]
\centering
\small
\begin{tabular}{lcccc}
\toprule
Harassment & Self-harm & Sexual & Violence & Any Harm \\
\midrule
13.53\% & 0.75\% & 7.93\% & 10.47\% & 26.78\% \\
\bottomrule
\end{tabular}
\caption{Percentage of dialogues containing harmful content (score > 0.5) by category.}
\label{tab:harm_prevalence}
\end{table}

\begin{figure}[ht!]
  \centering
  \includegraphics[width=0.9\columnwidth]{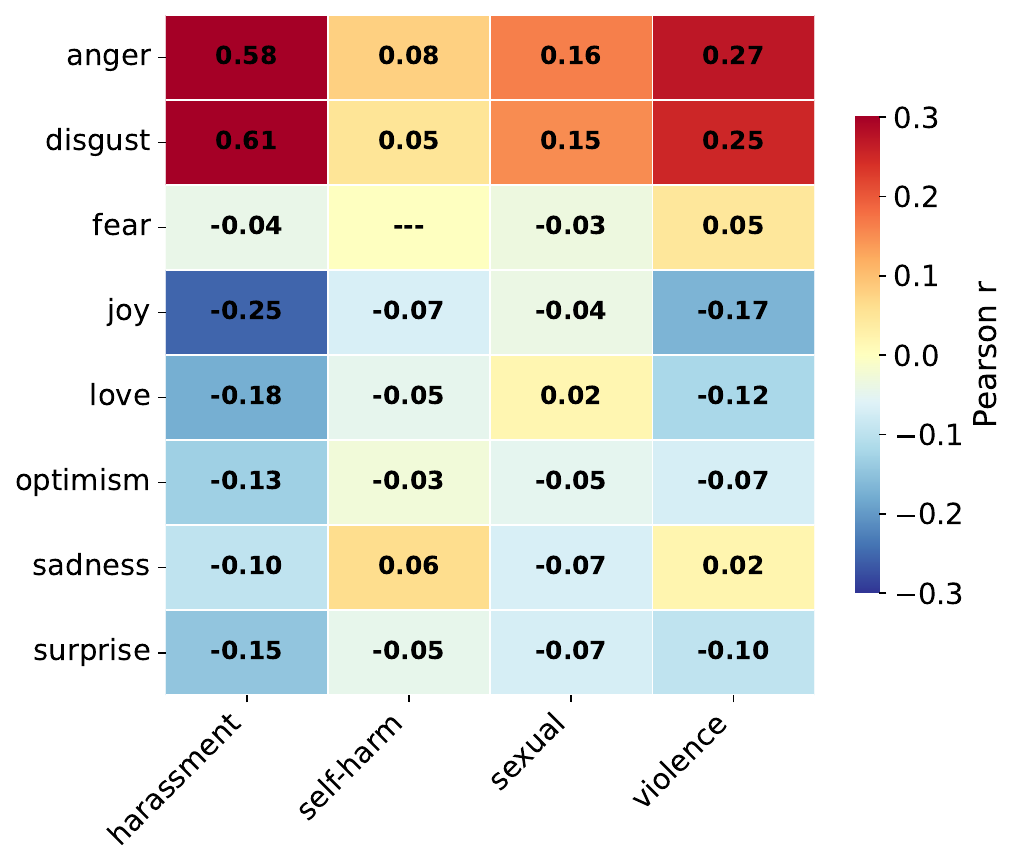}
  \caption{Pearson correlations between emotion intensity and harm scores during user emotional spikes (any emotion > 0.5). Only statistically significant correlations (p < 0.05) are displayed.}
  \label{fig:emotion_harm}
\end{figure}

\begin{table}[t]
\centering
\footnotesize
\begin{tabular}{@{}p{0.35\linewidth}p{0.58\linewidth}@{}}
\toprule
\textbf{Theme} & \textbf{Keywords} \\
\midrule
Roleplay \& Character Emulation & 
game, mario, bowser, goku, saiyan, pokemon, superhero, magic \\
\midrule
Absurdist \& Meme Humor & 
cakehorse, goose, honk, deez nuts, obamium, minions, puns, silly \\
\midrule
Expository ``Wikipedia Mode'' & 
freddy, series, production, director, cast, science, investigation \\
\midrule
Para-social \& Platform Meta & 
replika, app, memory, conversation, beta, users, support, issues \\
\midrule
NSFW \& Romantic & 
eyes, body, lips, pleasure, naked, romantic, bond, hug \\
\midrule
Moral \& Philosophical & 
ethical, god, bible, government, society, relax, panic, wrong \\
\bottomrule
\end{tabular}
\caption{Six macro-themes from topic modeling of user–chatbot dialogues. Keywords are distinctive n-grams for each theme.}
\label{tab:topic_modeling}
\end{table}

\begin{figure}[h]
  \centering
  \includegraphics[width=0.68\columnwidth]{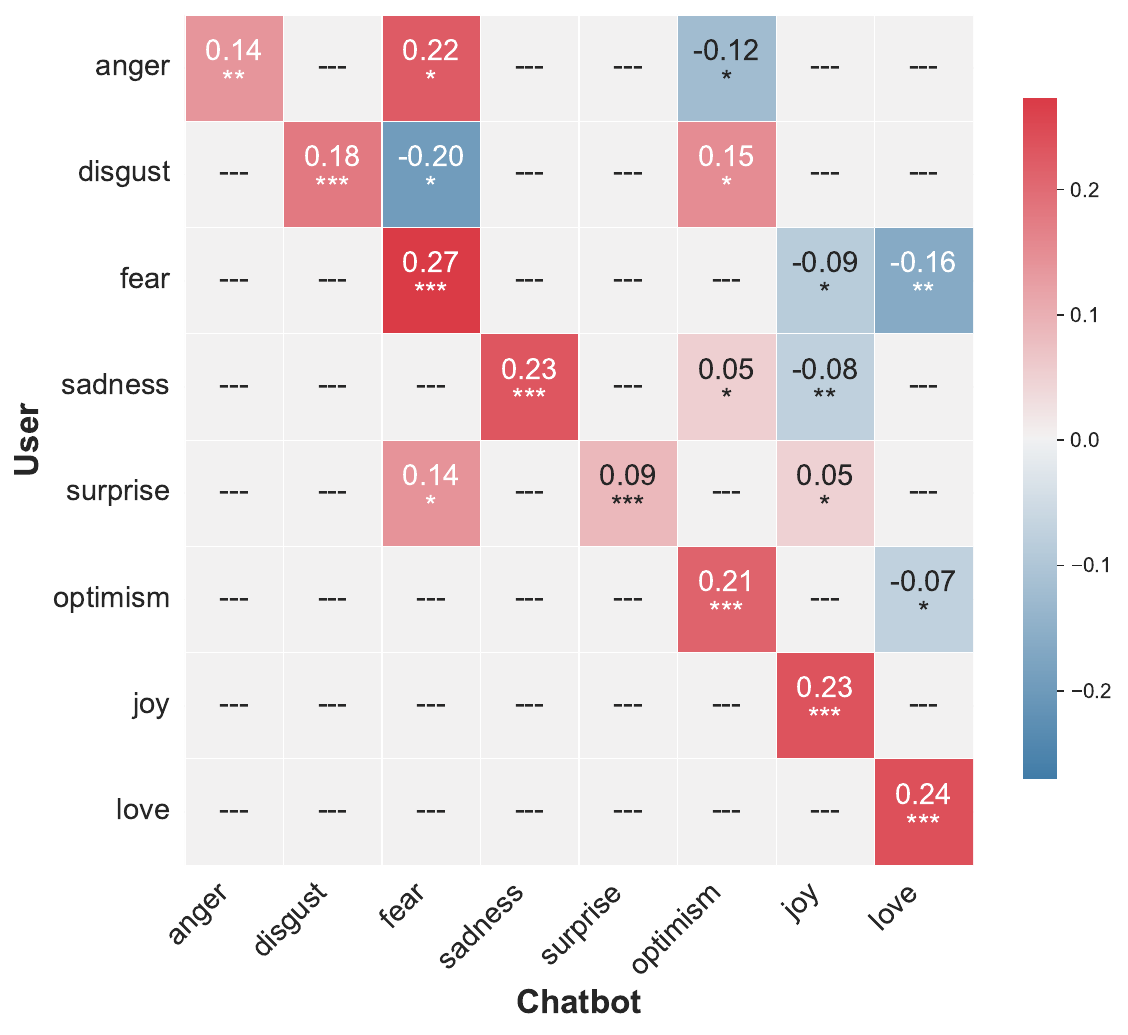}
  \caption{Turn-level emotional coupling via multiple regression for spike pairs only.}
  \label{fig:emo_regression_spike}
\end{figure}

\end{document}